\documentclass[12pt]{spieman} 
\pdfoutput=1
\usepackage[final]{graphicx}
\usepackage{times,bbm,amsmath,amssymb}
\usepackage{epsfig,color}
\usepackage{xcolor}
\usepackage{hyperref}
\usepackage{cleveref}
\usepackage{microtype}
\usepackage[symbol]{footmisc}

\usepackage{float,siunitx}
\usepackage[caption = false]{subfig}

\usepackage[greek,english]{babel}
\usepackage{thumbpdf,enumerate}
\usepackage{booktabs}
\usepackage{sidecap}
\usepackage[scaled=.8]{couriers}
\usepackage{multirow}
\usepackage{placeins}
\usepackage{relsize}
\usepackage{pst-grad,bm}
\usepackage{epigraph}
\usepackage{longtable}
\usepackage{ulem} 
\usepackage{acronym}
\usepackage{physics}
\usepackage{easyReview}

\usepackage{hyperref}

\newcommand{\bs}[1]{\boldsymbol{#1}}

\newcommand{\RR}{{\mathbb{R}}}
\newcommand{\calU}{{\mathcal{U}}}
\newcommand{\calH}{{\mathcal{H}}}

\title{Dynamical learning of a photonics quantum state-engineering process} 

\author[a,*]{Alessia Suprano}
\author[a,*]{Danilo Zia}
\author[a]{Emanuele Polino}
\author[a]{ Taira Giordani} 
\author[b,c,d]{Luca  Innocenti} 
\author[c]{Alessandro Ferraro} 
\author[c]{ Mauro Paternostro} 
\author[a]{ Nicol\`o Spagnolo} 
\author[a]{Fabio Sciarrino }
\affil[a]{Dipartimento di Fisica, Sapienza Universit\`{a} di Roma, Piazzale Aldo Moro 5, I-00185 Roma, Italy}

\affil[b]{Department of Optics, Palack\'{y} University, 17. Listopadu 12, 771 46 Olomouc, Czech Republic}
\affil[c]{Centre for Theoretical Atomic, Molecular, and Optical Physics,
School of Mathematics and Physics, Queen's University Belfast, BT7 1NN Belfast, United Kingdom}
\affil[d]{Università degli Studi di Palermo, Dipartimento di Fisica e Chimica – Emilio Segrè, via Archirafi 36, I-90123 Palermo, Italy}

\vspace{10pt}

\begin{document}
\maketitle
\begin{abstract}
Experimentally engineering high-dimensional quantum states is a crucial task for several quantum information protocols. However, a high degree of precision in the characterization of experimental noisy apparatus is required to apply existing quantum state engineering protocols. This is often lacking in practical scenarios, affecting the quality of the engineered states. Here, we implement experimentally an automated adaptive optimization protocol to engineer photonic Orbital Angular Momentum (OAM) states. The protocol, given a target output state, performs an online estimation of the quality of the currently produced states, relying on output measurement statistics, and determines how to tune the experimental parameters to optimize the state generation. To achieve this, the algorithm needs not be imbued with a description of the generation apparatus itself. Rather, it operates in a fully black-box scenario, making the scheme applicable in a wide variety of circumstances. The handles controlled by the algorithm are the rotation angles of a series of waveplates and can be used to probabilistically generate arbitrary four-dimensional OAM states. We showcase our scheme on different target states both in classical and quantum regimes, and prove its robustness to external perturbations on the control parameters. This approach represents a powerful tool for automated optimizations of noisy experimental tasks for quantum information protocols and technologies.
\end{abstract}

\vspace{10 cm}
{\noindent \footnotesize\textbf{*}These two authors contributed equally. }
\maketitle 

\section{Introduction}
Quantum state engineering of high dimensional states is a pivotal task in quantum information science~\cite{bechmannpasquinucci2000quantum,Vertesi2010,Lanyon2009,ralph2007efficient}. However, many existing protocols are platform-dependent, and lack universality~\cite{Anderson2015,Rossi2009,Hofheinz2009,Dada2011,Rosenblum2018,Heeres2017}.
Conversely, a scheme to engineer arbitrary quantum states, relying on Quantum Walk (QW) dynamics, was showcased in Ref.~\citeonline{giordani_2018}.
QWs are a particularly simple class of quantum dynamics which can be considered to generalize classical random walks~\cite{venegas-andraca2012quantum}. QWs have been implemented in experimental platforms ranging from trapped ions~\cite{schmitz2009quantum,zhringer2010realization} and atoms~\cite{karski2009quantum} to photonics circuits~\cite{sansoni2012quantum,crespi2013anderson,cardano2015quantum, caruso2016maze,kitagawa2012bound,qiang2016efficient,owens2011twophoton,boutari2016large}.
In particular, engineering of arbitrary qudit states has been experimentally demonstrated with QWs in the Orbital Angular Momentum (OAM) and polarization degrees of freedom of light~\cite{Innocenti2017,giordani_2018,suprano2021enhanced}.

In the paraxial approximation, the angular momentum of light can be decomposed in spin angular momentum, also referred to as polarization in this context, and OAM,
which is related to the spatial transverse structure of the electromagnetic field~\cite{allen_0AM_1992, yao2011orbital, piccirillo2013orbital}. 
In the classical regime, OAM finds application in particle trapping \cite{Zhan}, microscopy \cite{furhapter2005spiral,tamburini2006overcoming}, metrology \cite{lavery2013detection}, imaging \cite{Torner:05, Simon2012, Uribe-Patarroyo2013} and communication \cite{willner2015optical, bozinovic2013terabitscale, malik2012influence,baghdady2016multi,wang2016advances}.
On the other hand, in the quantum regime, OAM provides a high-dimensional degree of freedom, useful for example to encode large amounts of information in single-photon states.
Applications include quantum communication \cite{cozzolino_rev,Wang2015,krenn2015twisted,Malik2016,Sit17}, computing \cite{bartlett2002quantum, ralph2007efficient,Lanyon2009}, simulation \cite{cardano2016statistical,Buluta2009} and cryptography \cite{Mirhosseini_2015,Bouchard_18}.

Optimization algorithms have been proven to be useful tools in tasks such as detection of qudit states~\cite{li_2017} and quantum state engineering~\cite{Arrazola_2019, Mackeprang_2020}. 
Machine learning and genetic algorithms have also found many uses in photonics \cite{ma2021deep,wiecha2021deep}, including the use of generative models~\cite{Benedetti2019},
quantum state reconstruction~\cite{Yu2019,giordani2020machine}, automated design of experimental platforms~\cite{Melnikov,ren2021genetic,O'Driscoll2019}, quantum state and gate engineering~%
\cite{ Mackeprang_2020, Arrazola_2019,lumino2018experimental,santagati2018witnessing,wang2017experimental,rambhatla2020adaptive}, and the study of Bell nonlocality~\cite{Melnikov2020Bell,poderini2021ab,bharti2020machine}. Moreover, genetic algorithms have been employed to design adaptive spatial mode sorters using random scattering processes \cite{Fickler2017}. 
Between these types of algorithms, those based on a black box approach have the advantage that they do not rely on specific knowledge of the underlying experimental apparatus. These algorithms are built to tune a set of control parameters based on the information acquired from their environment, without having a notion of what the parameters represent 
in the experimental platform. This makes such approaches flexible and easier to apply in several scenarios.


In this paper, we showcase the use of \textit{RBFOpt}~\cite{Costa2018RBFopt, nannicini2021implementation}, a gradient-free global optimization algorithm based on radial basis functions~\cite{Powell1992, Powell1994, buhmann_2003}, 
to learn how to engineer specific quantum states by efficiently tuning the parameters of a given experimental apparatus.
The algorithm seeks to optimize a \textit{cost function} $C(\bs\Theta)$, with $\bs\Theta$ a set of real parameters determining the state produced by the apparatus.
As cost function $C(\bs\Theta)$, we use the quantum state fidelity between target and current state, as estimated from a given finite number of measurement samples. This makes the cost function inherently stochastic, and thus its optimization potentially more complex. 
As shown in Refs.~\citeonline{Costa2018RBFopt,nannicini2021implementation}, RBFOpt is particularly suited to optimize problems with few parameters, with a focus for operating regimes where only a small number of function evaluations is allowed. This is fully apt to our scenario, where functions evaluations involve the generation and measurement of photonic states, and are thus relatively costly.


We apply the proposed protocol to an experimental apparatus implementing discrete-time one-dimensional QWs in the OAM and polarization degrees of freedom of light, using a platform based on polarization-controlling waveplates and \textit{q-plates} \cite{marrucci-2006spin-to-orbital}: devices able to couple polarization and OAM degrees of freedom.
This platform was shown to be able to engineer arbitrary target quantum states~\cite{Innocenti2017,giordani_2018}.
Such approach, however, requires full knowledge of the inner workings of the underlying experimental apparatus. This feature makes it harder to flexibly adapt a protocol to the perturbations arising in realistic noisy conditions.
On the other hand, an adaptive algorithm operating in a black-box scenario, capable of finding the ideal control parameters independently of the physical substratum it operates in, is intrinsically more resilient to varying environmental and experimental circumstances.
To ensure the performance of our protocol is mostly independent on the specific task to which we apply it here, we avoided fine-tuning of the associated hyper-parameters, using the default values presented in Refs.~\citeonline{githubRBFOpt,RBFOptdoc}.
To further verify the resilience of the learning process, we also performed numerical simulations introducing some noise.



In~\cref{sec:1} we introduce the general optimization framework and the quantum walk model underlying our experimental architecture, and showcase the performance of the RBFOpt algorithm in numerical simulations with noise that mimics the experimental conditions.
In~\cref{sec:2} we describe the experimental platform and report how our optimization pipeline fares when operating directly on the experimental data.
In~\cref{sec:3} we analyze the performance of the protocol when applied to recover the optimal control parameters following sudden changes due to possible external perturbations, in order to probe its flexibility under different scenarios.
Finally, in~\cref{sec:4}, we summarize the results and lay our conclusions.
\begin{figure*}[h]
\includegraphics[width=1\textwidth]{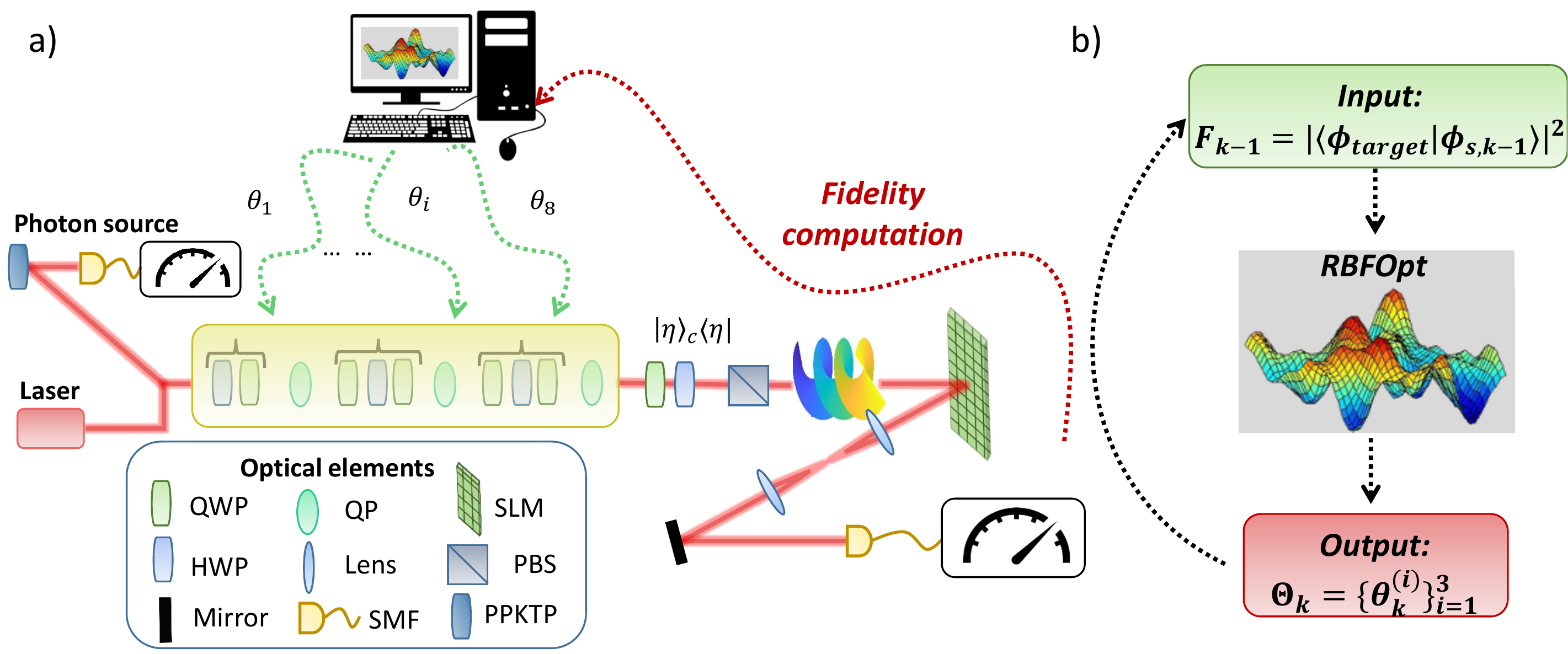}
\caption{\textbf{ Experimental Apparatus: 
a)} The engineering protocol has been tested experimentally in a three-step discrete-time QW encoded in OAM of light with both single-photon inputs and classical continuous wave laser light (CNI laser PSU-III-FDA) with a wavelength of $808$ nm. The single photon states are generated through a type-II spontaneous parametric down-conversion process in a Periodically-Poled KTP (PPKTP) crystal. The input state is characterized by a horizontal polarization and OAM eigenvalue $m=0$. Each step of the QW is made by a coin operator, implemented through a set of waveplates (QWP-HWP-QWP) and the shift operator, realized by a q-plate (QP).
To obtain the desired state in the OAM space a suitable projection in the polarization space is performed through a quarter-waveplate, a half-waveplate and a polarizing beam-splitter. The measurement station of the OAM-state is composed by a SLM followed by a single-mode fiber and the coupled signal is measured through a power meter, in the classical regime, or an Avalanche Photodiode detector, in the quantum one. In particular, in quantum optimizations pairs of photons are generated and heralded detection is performed computing the two-fold coincidences between detectors clicks from the QW evolved photon and the trigger one. The RBFOpt ignores the features of the experimental implementation that it is seen as a black box. 
The algorithm has access only to the $\Theta$ parameters of the coin operators and to the computed fidelity. \textbf{b)} During the iterations of the algorithm, the RBFOpt samples the black-box function to construct a surrogate model that is employed in the 
optimization. In the $k'$th iteration, the algorithm receives as input the fidelity computed in the previous iteration and uses it to improve the surrogate modeling. Moreover, the new parameters $\Theta_k$ are computed based on the optimization process. This procedure is repeated for each iteration of the algorithm.   }
\label{expsch}
\end{figure*}

\section{Quantum state engineering process as a black-box and simulated optimization}
\label{sec:1}

In order to study the effects of noise on the RBFOpt algorithm and its feasibility to engineer target quantum states, we apply it to numerically simulated data reproducing the most likely sources of noise in our experimental apparatus. We study, in particular, the effects of Binomial and Poissonian fluctuations on the cost function used by the algorithm.

Generating arbitrary qudit states is a pivotal and ubiquitous task in quantum information science and quantum technologies, with applications ranging from quantum communications~\cite{bechmannpasquinucci2000quantum, fitzi2001quantum, cerf2002security, bru2002optimal, acin2003security, langford2004measuring} to quantum computation~\cite{bartlett2002quantum, ralph2007efficient, Lanyon2009, campbell2012magicstate}.
The general quantum state engineering scenario we consider can be modeled with a parametrized unitary operation $\calU(\Theta)$, for some set of real parameters $\Theta\in\RR^N$.
Given a pair of initial and target states $\ket{\phi_{\rm in}}$ and $\ket{\phi_{\rm target}}$, the state engineering task consists of finding values $\Theta^\star\in\RR^N$ such that $\calU(\Theta^\star)\ket{\phi_{\rm in}}=\ket{\phi_{\rm target}}$.

%
To achieve this, we employ a numerical optimization algorithm to minimize the \textit{cost function} $C(\Theta)\equiv 1-F(\Theta)$, where $F(\Theta)\equiv\lvert\mel{\phi_{\rm target}}{\,\calU(\Theta)}{\phi_{\rm in}}\rvert^2$ is the fidelity between current and target states.
The optimization is performed in a fully black-box scenario, meaning we want the optimization procedure to be independent on the specifics of the particular optimization task. In particular, the optimization algorithm can control and optimize only the generation parameters $\Theta$ even if it has no knowledge about both generation of the output state ${\,\calU(\Theta)}\vert \phi_{\rm in} \rangle$ and computation of the cost function $C(\Theta)$. 
More specifically, we use {RBFOpt}~\cite{Costa2018RBFopt, nannicini2021implementation}, which works by building an approximated model of the objective function --- referred to as \textit{surrogate model} in this context --- using a set of Radial Basis Functions (RBFs).
RBFs are real-valued functions $\phi_{\bf p}$ that depend only on the distance from some fixed point:  $\phi_{\bf p}(\mathbf{x})=\phi(\norm{ \mathbf{x}-\mathbf{p}})$ for some $\phi$.
The goal of the surrogate model used in RBFOpt is to optimally exploit the information collected on the objective function from a limited number of function evaluations.
Based on the current estimation of the surrogate model, the algorithm selects new values of the control parameters to improve its current estimation of the model (see~\cref{app:A} for further details).
This algorithm is an extension of RBF algorithms~\cite{Gutmann2001RBF,MSRSM, Powell1992, Powell1994, buhmann_2003} whose performances are enhanced by providing an improved procedure to find an optimal surrogate model. A comparison of its performances with basic gradient-free algorithms is proposed in Appendix \ref{app:C}.

In our case, $\calU(\Theta)$ is the evolution corresponding to a one-dimensional discrete-time QW with time-dependent coin operations.
In this model, one considers states in a bipartite space $\calH_W\otimes\calH_C$, with $\calH_W$ a high-dimensional Hilbert space encoding the possible states of the \textit{walker} degree of freedom, and $\calH_C$ a two-dimensional space accommodating the \textit{coin} degree of freedom.
The dynamics is defined as a sequence of iterations, where each iteration is composed of a \textit{coin operation} $ \hat{C}(\bs\theta)$ followed by a \textit{controlled-shift operation} $\hat S$. To simulate the experimental conditions, the operators are defined as:
\begin{equation}
\begin{gathered}
    \hat{C}(\bs\theta)=
    \begin{pmatrix}
    e ^{-i\beta} \cos{\eta} & (\cos{\mu}+i\sin{\mu})\sin{\eta}, \\
    (-\cos{\mu}+i\sin{\mu})\sin{\eta} & e ^{i\beta} \cos{\eta} 
    \end{pmatrix}
    ,\\
    \hat{S} =
    \sum_k |k-1\rangle \langle k|_w\otimes |{\downarrow}\rangle \langle {\uparrow}|_c+ |k+1\rangle \langle k|_w\otimes |{\uparrow}\rangle \langle{\downarrow}|_c
    ,
\end{gathered}
\end{equation}
where $\beta\equiv\theta_1-\theta_3$, $\eta\equiv\theta_1-2\theta_2+\theta_3$, $\mu\equiv\theta_1+\theta_3$, and $\bs\theta\equiv(\theta_1,\theta_2,\theta_3)$ are the control parameters tuned by the algorithm.
This parametrization arises from the sequence of three polarization waveplates used to implement each coin operation. The case in which there are only two waveplates, as in the first step (\cref{expsch}), it is simply obtained from this putting $\theta_1=0$, and optimizing the values of $\theta_2$ and $\theta_3$.
Denoting with $\bs\theta^{(i)}\equiv(\theta^{(i)}_1,\theta^{(i)}_2,\theta^{(i)}_3)$ the free parameters characterising the coin operation at the $i$-th step, the full set of parameters characterising an $n$-step QW dynamics is then $\Theta=(\bs\theta^{(1)},...,\bs\theta^{(n)})\in\RR^{3n}$.
%
%
The overall evolution operator corresponding to $n$ steps is then $\calU(\Theta) \equiv \prod_{i=1}^n \hat S \hat{C}(\bs\theta^{(i)})$.
Following the engineering protocol presented in~\cite{Innocenti2017,giordani_2018}, we project the coin degree of freedom at the end of the evolution, so that our target state is some $\ket{\phi_{\rm target}}\in\calH_{W}$.

We apply RBFOpt to optimize a three-step QW, where in the first iteration only two free parameters are used. This corresponds to a total of $8$ control parameters: $\Theta=(\bs\theta^{(i)})_{i=1}^{3}$ with $\bs\theta^{(1)}\equiv(0,\theta^{(1)}_2,\theta^{(1)}_3)$. Importantly, the algorithm does not use the information of the correct model $\calU(\Theta)$ of the evolution. This feature permits to use the present approach in conditions where a model of the experimental setup and noise processes is lacking.

In order to simulate the experimental calculation of the fidelity of a given target state, an orthonormal basis $\{\vert \psi_j \rangle \}_{j=1}^d$, where $d$ is the dimension of the target state and $\vert \psi_1 \rangle=\vert \phi_{\rm target} \rangle$, is built through the Gram-Schmidt algorithm.
This approach to estimate the cost function is used to simulate the experimental statistics collection process.
We furthermore consider both Poissonian ($\mathcal{P}({\lambda})$) and a Binomial ($\mathcal{B}(N,p)$) fluctuations.
Poissonian fluctuations are introduced to take into account laser oscillations, while Binomial fluctuations reflect the probabilistic nature of the measurement setup. 

The number of events of the Binomial distribution $N$ is extracted from a Poissonian distribution with a parameter $\lambda = 10^4$, while the probability $p$ is equal to the fidelity between the state proposed by the algorithm, in the $k'$th iteration, and the specific element of the basis. 
Therefore, for each element of the orthonormal basis the number of detected events is extracted from the Binomial distribution. The noisy fidelity between the proposed state and the target state is then calculated as the ratio between the counts for the element $\vert \psi_1 \rangle$ and the total number of counts.

\begin{figure}[h]
\begin{center}
\includegraphics[width=0.5\columnwidth]{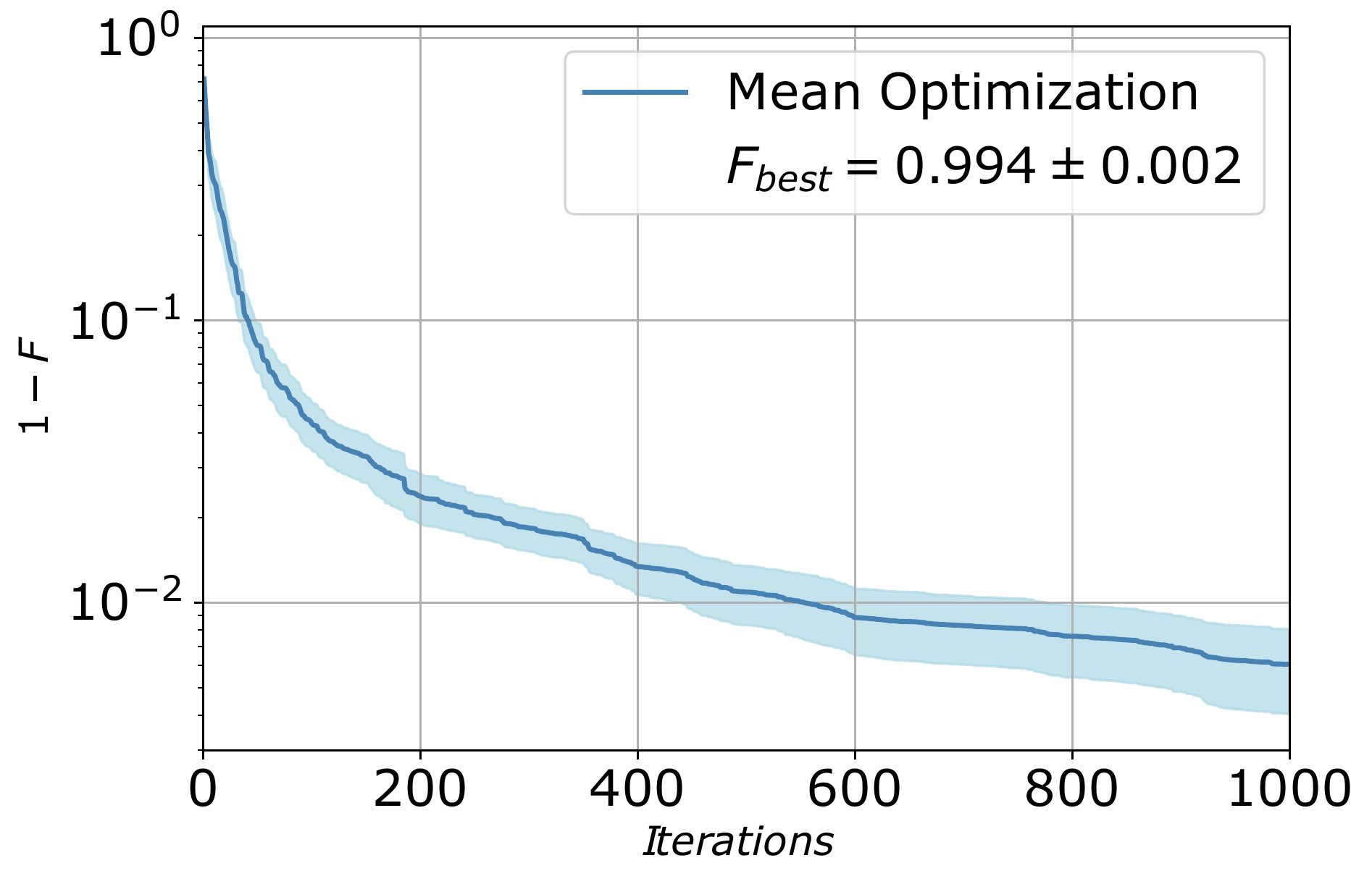}
\caption{\textbf{ Simulated Optimization.} Infidelity $1-F$ obtained at different stages of the optimization. We test the algorithm on $10$ random target states, repeating the optimization $10$ time for each. The reported results are obtained as the mean over the average behavior for each of the $10$ states. 
The highest average fidelity obtained is $0.994 \pm 0.002$. The shaded area represents the standard deviation of the mean.
}
\label{simres}
\end{center}

\end{figure}

We apply the optimization protocol to $10$ random four-dimensional target states, repeating the optimization $10$ times for each state.
In~\cref{simres} we show the value of the cost function --- \textit{i.e.} the infidelity between current and target states --- obtained at different stages of the algorithm, up to the fixed maximum number of $1000$ iterations. For each iteration number, we report the infidelity obtained as the mean over the average behavior of each of the $10$ states. The obtained trend demonstrates that, also in noisy conditions, the algorithm manages to minimize the function, and promising results are obtained.
Moreover, we also investigate the scalability of the proposed approach when the number of parameters increases. In particular, we simulated QWs of up to 17 steps (50 parameters) and observed in the investigated regime a linear increase in the mean number of iterations needed to achieve a fidelity value of at least $98 \%$. Further details are reported in Appendix B. 

\section{Experimental Dynamical Learning}\label{sec:2}

The capability to manipulate the OAM of light enabled effectiveness experimental implementations of high-dimensional discrete-time QWs.
Therefore, to test experimentally the optimization procedure, we exploit a setup based on the scheme proposed in Ref. \citeonline{giordani_2018}. In particular, we implemented three steps of a discrete-time QW encoding the coin state in the photon polarization, and the walker in the OAM degree of freedom.
At each iteration, the coin operation is implemented as a set of polarization \textit{waveplates}, while the controlled-shift via a \textit{q-plate}, a device that acts on the OAM conditionally on the polarization state of the light~\cite{marrucci-2006spin-to-orbital}:
\begin{equation}
    \hat{Q}=\sum_m |m-1\rangle \langle m|\otimes |L\rangle \langle R|+ |m+1\rangle \langle m|\otimes |R\rangle \langle L|
\end{equation}
 where $m$ is the OAM value and, $R$ and $L$ are the right and left circular polarizations, respectively.
We implement arbitrary coin operations using two quarter-waveplates interspaced (QWP) with a half-waveplate (HWP).
The output OAM state is then obtained performing a suitable projection on the polarization.
This is implemented with a set of waveplates followed by a polarizing beam-splitter (\textit{c.f.}~\cref{expsch}-a).

To measure the fidelity of the output states, we use a measurement apparatus composed of a Spatial Light Modulator (SLM) \cite{bolduc2013holo,forbes2016creation} and a single-mode fiber.
Since the SLM modulates the beam shape through computer-generated holograms, the operation of this measurement station is equivalent to a projective measurement on the state encoded in the employed hologram.
To characterize an incident beam, we thus display on the SLM the hologram corresponding to each element of an orthonormal basis, obtaining the corresponding fidelities. The optimization speed is mainly limited by the measurement process since significant statistics has to be collected for each projected hologram. Therefore, the use of algorithms able to limit the objective function evaluations, such as those based on the building of a surrogate model, is preferable. 

\begin{figure*}[h]
\begin{center}
\includegraphics[width=\textwidth]{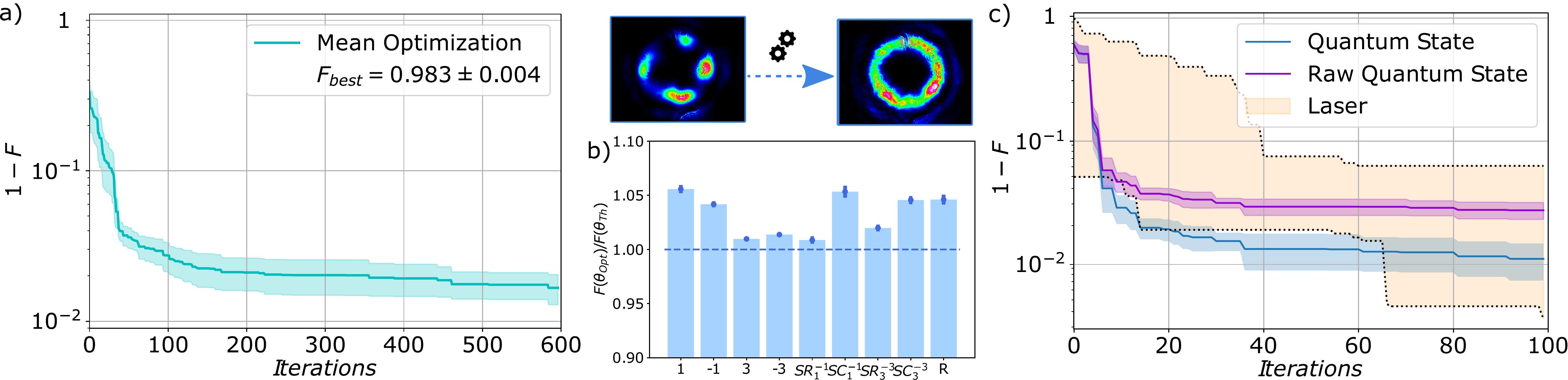}
\caption{\textbf{ Experimental Results:} \textbf{a)}  Minimization of the quantity $1-F$ averaged over the algorithm performances for different experimental states. The mean maximum value reached is $0.983 \pm 0.004$. \textbf{b)} Ratio between the maximum experimental values of the fidelities resulted after the optimization $F(\Theta_{Opt})$ and the fidelities measured with the theoretical parameters $F(\Theta_{Th})$. For each engineered state, the ratio is higher or compatible with the value 1 highlighted by the dashed line. This confirms that the adopted algorithm can reach performances compatible or even superior with respect to 
the one obtained with the direct method presented in Ref. \citeonline{Innocenti2017} that consider ideal experimental platforms. In this sense, the algorithm can take into account and compensate for the experimental imperfections. All the error bars reported are due to laser fluctuations affecting each measurement and are estimated through a Monte Carlo approach. \textbf{c)} Comparison between the performances reached in $100$ iterations using classical or single photon input states. In yellow is reported the area between the best and worst optimization performed in the classical case. The blue and violet curves are associated to the minimization of the quantity $1-F$ averaged over $5$ different optimizations for the state $SR_1^{-1}$ engineered in the quantum domain. In particular, in violet are shown the raw data, while in blue the data after accidental counts subtraction.
}
\label{expres}
\end{center}

\end{figure*}

The computed fidelities are then fed to the RBFOpt algorithm to tune the waveplate parameters $\Theta$. To achieve this, the algorithm does not require knowledge on the final target state or on the generation and measurement functioning, as shown in~\cref{expsch}-b.
However, since the algorithm has no control over the measurement station, the parameters of the latter have been fine-tuned a priori and we are confident of the correctness of this step. Therefore, through a dynamic control of the waveplates orientation, the algorithm is able to optimize the fidelity value in real time. 

To showcase the efficiency of the protocol on our experimental platform, we applied it to engineer different kinds of target states in both classical and quantum regime.
In ~\cref{expres} we show the results of running the optimization algorithm on 9 different classical states. In particular, we focus our analysis on the elements of the computational basis $\vert m \rangle$ with $m \in \{-1,1,-3,3\}$ and on the balanced superposition of two OAM values. We considered both real $SR_{m_1}^{m_2}=\frac{\vert m_1 \rangle - \vert m_2 \rangle}{\sqrt{2}}$ and complex superpositions $SC_{m_1}^{m_2}=\frac{\vert m_1 \rangle -i \vert m_2 \rangle}{\sqrt{2}}$, where $m_1,m_2 \in \{-1,1,-3,3\}$ with $\vert m_1 \vert=\vert m_2 \vert$. Moreover, to verify the efficiency of the protocol we optimize the engineering of a randomly extracted state ($R$) in the four-dimensional Hilbert space with no zero coefficients corresponding to each basis element.
As shown in the panel a of~\cref{expres}, optimal average values are obtained in $600$ algorithm iterations. In particular, the reported infidelity $1-F$ is computed averaging over all the experimentally engineered states, and the minimization is compatible with the numerical results reported in~\cref{simres}.
In~\cref{expres}-b we report, for each engineered state, the ratio between the fidelities found by the RBFOpt algorithm, and those found using the method presented in~\cite{Innocenti2017} to find the optimal values of the parameters. 
Indeed, as demonstrated in Ref. \citeonline{Innocenti2017}, it is possible to find coin parameters resulting in an arbitrary target state --- albeit possibly with different projection probabilities --- regardless of the experimental conditions.
We find the fidelities reached by RBFOpt to be always higher than the ones computed using the direct method presented in~\cite{Innocenti2017}.
This is due to the dynamical learning algorithm we employ, which shows higher performances in compensating experimental imperfections.
This showcases the advantanges of real-time optimization algorithms for quantum state engineering in realistic scenarios. Notably, we extended the experimental demonstration of the protocol also in quantum regime of single photon states. We showcased the engineering of the superposition state $SR_{1}^{-1}$ and we repeated $5$ times the optimization, considering only 100 iterations. No differences are 
    expected between the employment of laser and single photon states. In \cref{expres} c we compared the two performances 
    and observed a good agreement between the 
    approaches. In particular, we reported the optimization curves obtained in the quantum regime plotting the raw data, corresponding to $\sim 4000$ Hz coincidences, and by subtracting the accidental counts. This allows us to distinguish the contribution to the cost function given by either the engineering or the measurement system. The corresponding maximum mean fidelities are $F=0.972 \pm 0.003$ and $F=0.989 \pm 0.003$, respectively. In conclusion, since very high fidelities are reached in only $100$ steps, the proposed approach can be efficiently applied to quantum situations.



\begin{figure*}
\begin{center}
\includegraphics[width=.90\textwidth]{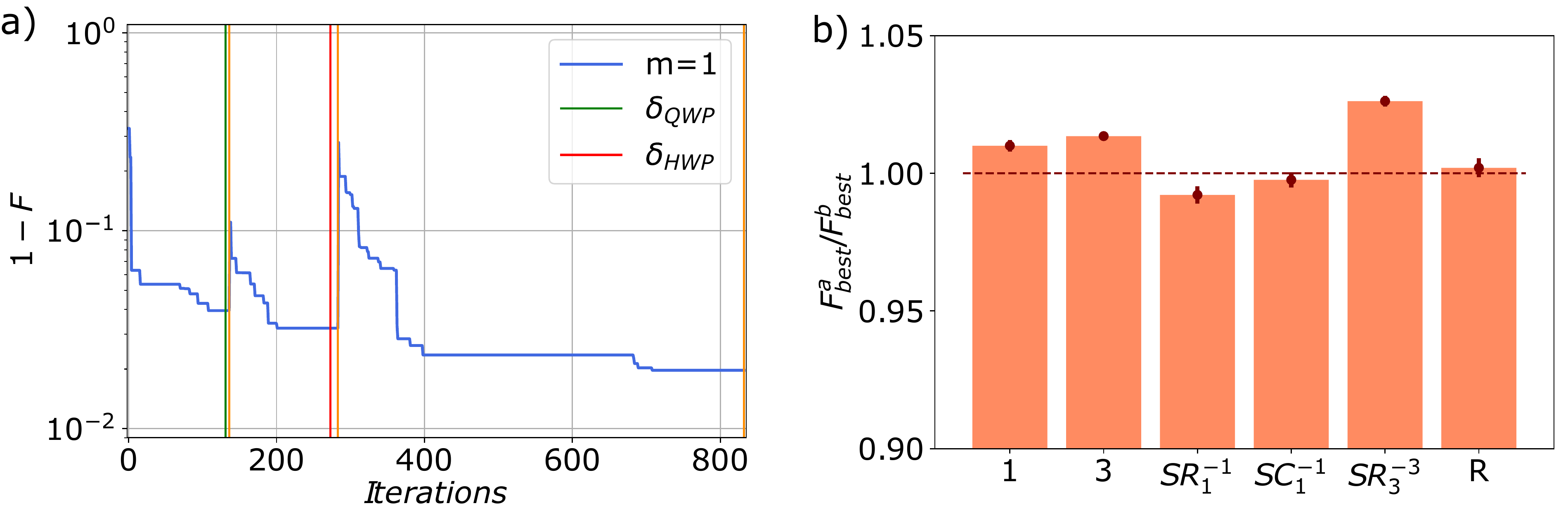}
\caption{\textbf{ Experimental Perturbation Results: a)} Optimization under external perturbation of the quantity $1-F$ for the state $\ket{1}$. The iterations in which a perturbation $\delta$ occurs are highlighted by a vertical red line (second step HWP) or by a vertical green line (third step QWP), a vertical orange line highlights the iteration in which the algorithm is restarted. \textbf{b)} Mean ratio between the best value obtained for the fidelity after ($F_{best}^{a}$) and before ($F_{best}^{b}$) the perturbation for the different engineered states. The ratio is close to or higher than 1 for all of them, this showcases that the algorithm is able to re-obtain and eventually improve the best value sampled before the perturbation. All the error bars reported are due to laser fluctuations affecting each measurement and are estimated through a Monte Carlo approach.}
\label{kickres}
\end{center}

\end{figure*}

\section{Dynamical Learning protocol with external perturbations}\label{sec:3}

In realistic conditions, noise is unavoidable,
which makes the capability of an algorithm to adapt to real-world perturbations pivotal.
To test the robustness of RBFOpt, we have thus added external perturbations to the experimental setup. In particular, we consider a scenario where a sudden perturbation on the parameters is introduced.
The algorithm is then tasked with finding again the optimal parameters required to engineer the target state.
We assess the performances of the algorithm throughout the optimization, to determine  whether a perturbation occurred, and thus the control parameters need to be re-optimized.
More specifically:

\begin{enumerate}
\item Every $10$ iterations, we used the optimal parameters found by the algorithm up to that time $\Theta_{\rm best}$ to obtain a new estimate of the cost function $C_{\rm new}(\Theta_{\rm best})$.
\item To spot if a perturbation occurred, we compared the new value with the one obtained during the algorithm evolution $C_{\rm sampled}(\Theta_{\rm best})$. So chosen a threshold $t$, we proceed as follows:
\begin{enumerate}
\item If $C_{\rm new}(\Theta_{\rm best}) \leq C_{\rm sampled}(\Theta_{\rm best}) + t$, the optimization is continued.
\item If $C_{\rm new}(\Theta_{\rm best}) > C_{\rm sampled}(\Theta_{\rm best}) + t$, the algorithm is restarted.
\end{enumerate}
\end{enumerate}
Therefore, within this approach the surrogate model is discarded and rebuilt from scratch every time the quantity of interest is deteriorating. We performed this check every 10 algorithm iterations in order to have a quick response to perturbations without excessively increase the optimization time. Indeed, each function evaluation consists in a time-consuming projective measurement with the SLM. For each engineered state the value of the threshold was fixed analysing the fluctuations in the value of the measured fidelity $F$, these values are reported in Table \ref{Tab:KickPar}.


The considered perturbations
act on the HWP of the second step and on the first QWP of the third step, this disturbance consists in a permanent offset in the waveplates rotation of a quantity $\delta$. In particular, at each iteration and with probability $q$ the orientation of the waveplates optical axis is changed by the addition of an angle sampled from a normal distribution with mean $\mu = - \ang{30}$ and standard deviation $\sigma = \ang{5}$ ($\mathcal{N}(-\ang{30}, \ang{5})$). We investigated the algorithm response using elements of the computational basis, balanced superpositions of such elements and a random state. In these cases several values for the parameter $q$ are used. The engineered states and the probability $q$ used for them are reported in Table \ref{Tab:KickPar}.
\begingroup
\setlength{\tabcolsep}{10.5 pt}
\renewcommand{\arraystretch}{0.3} 
\begin{table}[htp]
\centering
\begin{tabular}{ 
  c
  c
  c  }
\toprule
Target & Perturbation & Restart \\
State& Probability & Threshold \\
\midrule
\;&\;&\;\\
$\ket{1}$ &  \;\;$0.0015$\;\; & $0.02$ \\
\;&\;&\;\\
$\ket{3}$ &  \;\;$0.0015$\;\; & $0.02$ \\
\;&\;&\;\\
$\frac{1}{\sqrt{2}}(\ket{-1}+\ket{1})$ & \;\;$0.008$\;\;  & $0.02$ \\
\;&\;&\;\\
$\frac{1}{\sqrt{2}}(\ket{-1}+i\ket{1})$ & \;\;$0.004$\;\;  & $0.02$ \\
\;&\;&\;\\
$\frac{1}{\sqrt{2}}(\ket{-3}+\ket{3})$ & \;\;$0.0015$\;\;  & $0.05$ \\
\;&\;&\;\\
Random & \;\;$0.0015$\;\;  & $0.02$ \\
\;&\;&\;\\
\bottomrule
\end{tabular}
\caption{The table shows the parameters used in the study of the optimization under perturbations for the engineered states. In the second column we report the values of the perturbation occurrence probability $q$, while in the third column we report the threshold values $t$ used for deciding the algorithm restart.}
\label{Tab:KickPar}
\end{table}
\endgroup

An example for the dynamics under perturbations is reported in \cref{kickres}(a), here the iteration in which a disturbance is introduced is highlighted by a vertical red or green line respectively for shift on the HWP or on the QWP. Instead, the restart of the algorithm is indicated with a vertical orange line. As shown, after the perturbation, the minimum found by the algorithm is no longer the optimal solution thus triggering a restart. The latter allows the algorithm to reach a new optimal solution in a different environmental condition.
Moreover, in panel b the mean ratio between the best fidelity found before and after a perturbation is reported for each analysed state.
Knowing that for each state more than one perturbation could be performed, the mean ratio is computed averaging over all of them. Here values close to or grater than 1 point out how, thanks to the restart, the algorithm is able to re-adapt its optimal solution and eventually improve the previously obtained fidelity.


\section{Conclusions}
\label{sec:4}

The black-box optimization paradigm we discussed is highly flexible, thus promising to be a powerful tool with the potential to be applicable to problems ranging from optimizations of quantum information platforms to the study of nonclassicality.

We have showcased how the RBFOpt global optimization algorithm allows to dynamically learn the quantum state generation process.
In particular, such approach enables the optimization of target states engineering  without having to devise \textit{ad hoc} platform-dependent protocols.
First of all, we dynamically tune the QW parameters in order to optimize the engineering of 9 different experimental states in the classical domain. The obtained results turned out to be comparable to the preliminary ones achieved in our numerical simulations. Moreover, the RBFOpt results in higher fidelities than those computed using the direct method of Ref.~\citeonline{Innocenti2017}. Therefore, the real-time optimization allows to take into account and compensate experimental imperfections. Moreover, we optimized an experimental state using a single photon source as input to prove the equivalence between the performances reached in the classical and quantum regime and extend the proposed approach. In order to carry out a complete analysis, and as the adaptation capability of an algorithm is pivotal in realistic condition, we simulated the effect of real-world perturbations. We have thus applied the optimization algorithm to different states while adding permanent offsets to the orientation of two waveplates in a probabilistic manner. The algorithm menages to adapt itself so as to reach fidelities comparable with those obtained before the perturbation. Our results proves the advantages of adopting real-time optimization algorithms for experimental quantum state engineering protocols. Therefore, practical
experimental quantum information experiments can benefit from our work, increasing the engineering performances and employing a real-time fine-tuning of the parameters. The proposed approach can be extended to many different tasks, as an example, by suitably modifying the cost function, it is possible to optimize not only the fidelity but also the success probability to generate a target state after the coin projection (see Ref.~\citeonline{giordani_2018,Innocenti2017} ). Moreover, since the algorithm does not require information on the function to be optimized and on the employed experimental platform, our scheme can find applications in different engineering protocols and quantum information tasks that make use of controllable devices parameters employing, in principle, arbitrary degree of freedoms.
Furthermore, going beyond the fully black-box paradigm, in principle, the approach can be exploited also for different protocols. For instance in the theoretical design of experiments, it could be used to optimize the number of quantum gates needed for a specific desired task. Moreover, it could be used also in the calibration of complex optical circuits that find applications in tasks like Boson Sampling ~\cite{zhong2020quantum,brod2019photonic,hoch2021boson,cimini2021calibration} and in the engineering of multiphoton quantum states ~\cite{pan2012multiphoton}. In this case, it would be crucial to tailor a suitable cost function.

\section*{Acknowledgments}
We acknowledge support from the European Union's Horizon 2020 research and innovation programme (Future and Emerging Technologies) through project TEQ (grant nr. 766900), QU-BOSS-ERC Advanced Grant (grant nr. 884676), the QUSHIP PRIN 2017 (grant nr. 2017SRNBRK), the DfE-SFI Investigator Programme (grant 15/IA/2864), COST Action CA15220, the Royal Society Wolfson Research Fellowship (RSWF\textbackslash R3\textbackslash183013), the Leverhulme Trust Research Project Grant (grant nr.~RGP-2018-266), the UK EPSRC (grant nr.~EP/T028106/1).

\appendix

\section{Description of the RBFOpt algorithm}
\label{app:A}

The RBFOpt optimization algorithm is based on the exploitation of a radial basis interpolant, called surrogate model \cite{Gutmann2001RBF,Costa2018RBFopt,nannicini2021implementation, MSRSM}. Given $k$ distinct parameter points $\Theta_{1},...,\Theta_{k} \in \Omega$, where $\Omega$ is a compact subset of $\mathbb{R}^{N}$, with corresponding cost function values $C(\Theta_{1}),...,C(\Theta_{k})$. The associated surrogate model $ s_{k}(\Theta)$ is defined as:
\begin{equation}
    s_{k}(\Theta) =\sum_{i=1}^{k} \lambda_{i} \phi(\norm{\Theta-\Theta_{i}}) + p(\Theta)
\label{eq:surrogate}
\end{equation}
where $\phi(.)$ is a radial basis function, $\lambda_{1},...,\lambda_{k} \in \mathbb{R}$ and $p(.)$ is a polynomial of degree $d$. This degree is selected based on the type of the RBF function used in the surrogate model. The possible RBF function choices and the degree of their associated polynomial are reported in Table \ref{Tab:RBF}. The hyperparameter $\gamma$ present in the expression of the radial basis functions is set to 0.1 by default \cite{RBFOptdoc,githubRBFOpt}. 
Moreover, the RBFOpt algorithm automatically selects the radial basis function that appears to be the most accurate in the description of the problem. This selection is made using a cross validation procedure, in which the performance of a surrogate model constructed with points $\left( \Theta_{i}, C(\Theta_{i})\right)$ for $i=1,....,k$ are tested at $\left(\Theta_{j}, C(\Theta_{j})\right)$ with $j \neq i$ \cite{Costa2018RBFopt,nannicini2021implementation}. 

The value of the parameters $\lambda_{i}$ with $i=1,...,k$ and the coefficients of the polynomial can be determined solving the following linear system \cite{Gutmann2001RBF,Costa2018RBFopt,nannicini2021implementation,MSRSM}:
\begin{equation}
\begin{cases}
     s_{k}(\Theta_{i}) = C(\Theta_{i}),\;\;\;i = 1,...,k\\
     \;\\
     \sum_{i=1}^{k}\lambda_{i}\hat{p}_{j}(\Theta_{i}) = 0,\;\;\;j = 1,...,\tilde{d} 
    \end{cases}
\label{eq:linear}
\end{equation}
where, called $\Pi_{d}$ the space of polynomials of degree less than or equal to $d$, $\tilde{d}$ is the dimension of $\Pi_{d}$ and $\hat{p}_{1},...,\hat{p}_{\tilde{d}}$ is a basis of the space.
\begingroup
\setlength{\tabcolsep}{16.5 pt}
\renewcommand{\arraystretch}{0.7} 
\begin{table}[htp]
\centering
\begin{tabular}{ 
  c 
  c  }
  \toprule
Radial Basis Function $\phi(x)$ & \;\;Polynomial degree $d$\\
\midrule
$x$ &   $0$ \\
$x^{3}$ &  $1$ \\
$\sqrt{x^{2}+\gamma^{2}}$ & $0$ \\
$x^{2}\log{x}$ &  $1$ \\
$e^{-\gamma x^{2}}$ &  $-1$ \\
\bottomrule
\end{tabular}
\caption{The RBFs exploited by the RBF algorithm and the degree of the polynomial used in the construction of the surrogate model \cite{Gutmann2001RBF,Costa2018RBFopt,nannicini2021implementation,MSRSM}. When $d=-1$ the polynomial is removed from the expression \eqref{eq:surrogate}.}
\label{Tab:RBF}
\end{table}
\endgroup

At the beginning of the optimization procedure, the surrogate model is constructed from a set of 
parameter points tunable in number and sampled using a latin hypercube design \cite{githubRBFOpt,RBFOptdoc}. After that, the interpolant is used to choose the next point in which compute the cost function. So, the evolution of the RBFOpt algorithm is  composed by the repetition of following steps (say $k$'th step):
\begin{enumerate}
    \item Compute the surrogate model $s_{k}(\Theta)$ from the data points $\left(\Theta_{i}, C(\Theta_{i})\right)$, with $i=1,...,k$, solving the linear system of Eq.(\ref{eq:linear})
    \item Use the surrogate model to choose the next point $\Theta_{k+1}$. In particular, the Metric Stochastic Response Surface Method (MSRSM) is applied \cite{Costa2018RBFopt,nannicini2021implementation,MSRSM}. Within this framework, the algorithm does a number of \textit{global steps} controlled by the hyperparameter \textit{num\_global\_searches} (default value $5$ \cite{RBFOptdoc,githubRBFOpt}) and a \textit{local step}. The latter gives as next point the one that minimizes the surrogate model.
    \item Evaluate the cost function at $\Theta_{k+1}$ and add $\left(\Theta_{k+1}, C(\Theta_{k+1})\right)$ to the data points.
    \item Decide if restart the model for lack of improvement. Specifically, if the algorithm doesn't find a new optimal solution after a number of evaluations defined by the hyperparameter $max\_stalled\_iterations$ (default value 100 \cite{RBFOptdoc,githubRBFOpt}), the actual surrogate model is discarded and the optimization procedure is restarted from scratch.

\end{enumerate}

Moreover, during the optimization the algorithm executes a \textit{refinement step}. The purpose of which is to improve the optimal solution doing a local search around it through a variation of a trust region method \cite{Costa2018RBFopt,nannicini2021implementation}. The refinement step is triggered at the end of point (3) with a frequency controlled by the hyper-parameter \textit{refinement\_frequency}, with default value equal to $3$\cite{RBFOptdoc,githubRBFOpt}.\\
Furthermore, in the study concerning the evolution under external perturbation we add, as explained in section \ref{sec:3}, a new condition for triggering a restart. Beyond the default one, we analyzed the deterioration of the optimal value founded for the cost function and decided if restart the optimization. This further check was done every $10$ iterations in order to have a faster response to perturbations without increasing excessively the number of function evaluations, that, experimentally, is expensive in time. 

\section{Scalability of the optimization approach}
\label{app:B}
In this section, we study the RBFOpt behaviour as the number of parameters of the objective function increases.
In particular, we simulated different experimental configurations with quantum walk steps ranging from 3 to 17 and thus considering up to 50 parameters. In fact, being $N_{steps}$ the number of steps and considering only two waveplates in the first coin, the number of parameters $N_{par}$ follows the relation:
\begin{equation}
    N_{par} = 3 \; N_{steps}-1
\end{equation}
For each case, we generated at random 50 target states and investigated the optimization procedure stopping the process when a fidelity of at least $98 \%$ was reached. In all the evolutions, we added the same Poissonian and Binomial noises described in the main text to the fidelity between the target state and the one proposed by the algorithm.

\begin{figure*}[htp]
\begin{center}
\includegraphics[width=0.5\textwidth]{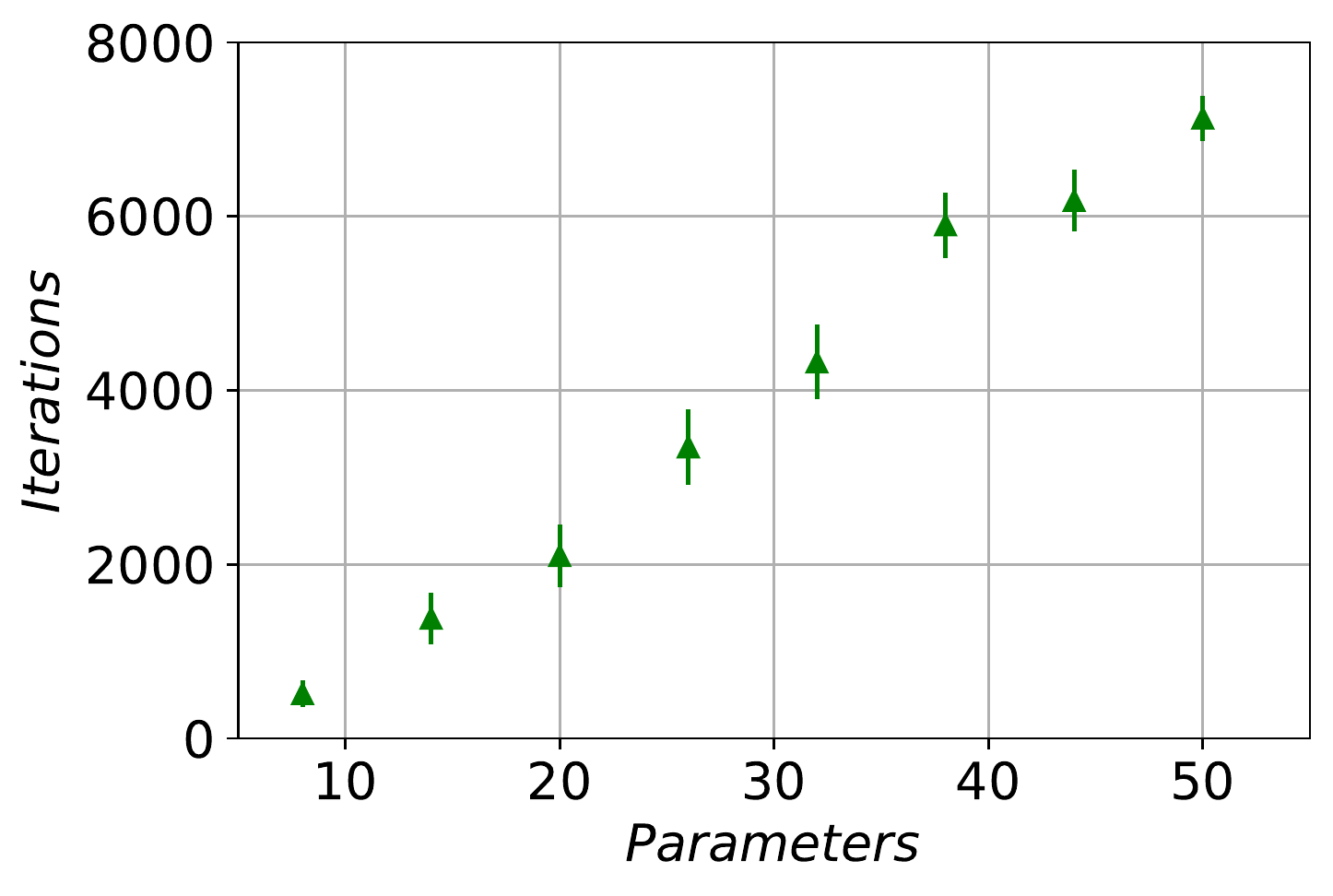}
\caption{\textbf{Scalability:} The plot shows the mean number of RBFOpt algorithm iterations as a function of the black-box problem parameters. Here, the optimization process is interrupted when a value of the fidelity between the target state and the one proposed by the algorithm of at least $98 \%$ is reached. For each configuration, the iterations values are obtained averaging over $50$ random target states and simulating experimental noise using Binomial and Poissonian distributions. The uncertainty associated with each point is provided by the standard deviation of the mean.}
\label{fig:Scalability}
\end{center}
\end{figure*}

The computational cost of performing a black-box optimization in high dimensional spaces can be extracted analysing how the mean number of iterations changes in relation to the number of parameters. The values obtained averaging over the 50 states considered in our study are reported in Fig. \ref{fig:Scalability} for each simulated configuration.
As can be seen from the plot, the RBFOpt algorithm appears to have a linear scaling over the parameters number when applied to our implementation. This theoretically showcases the effectiveness of the proposed approach for the  engineering of higher dimensional OAM states and similar behaviours are expected experimentally taking into account the devices response time and adapting  properly the related implementation. Finally, while similar behaviors are expected in the regime of few parameters, for higher orders of magnitude the time needed to perform an iteration step increases drastically. In such regimes, a more refined version of the algorithm might be useful to improve its efficiency.



\section{Comparison between RBFopt and basic algorithms}
\label{app:C}
In this section, we perform simulations to compare the RBFOpt algorithm with two basic gradient-free methods suitable to multi-parameters black-box optimization. In particular, we consider both non adaptive and adaptive approaches.

\begin{figure*}[h]
\begin{center}
\includegraphics[width=0.5\textwidth]{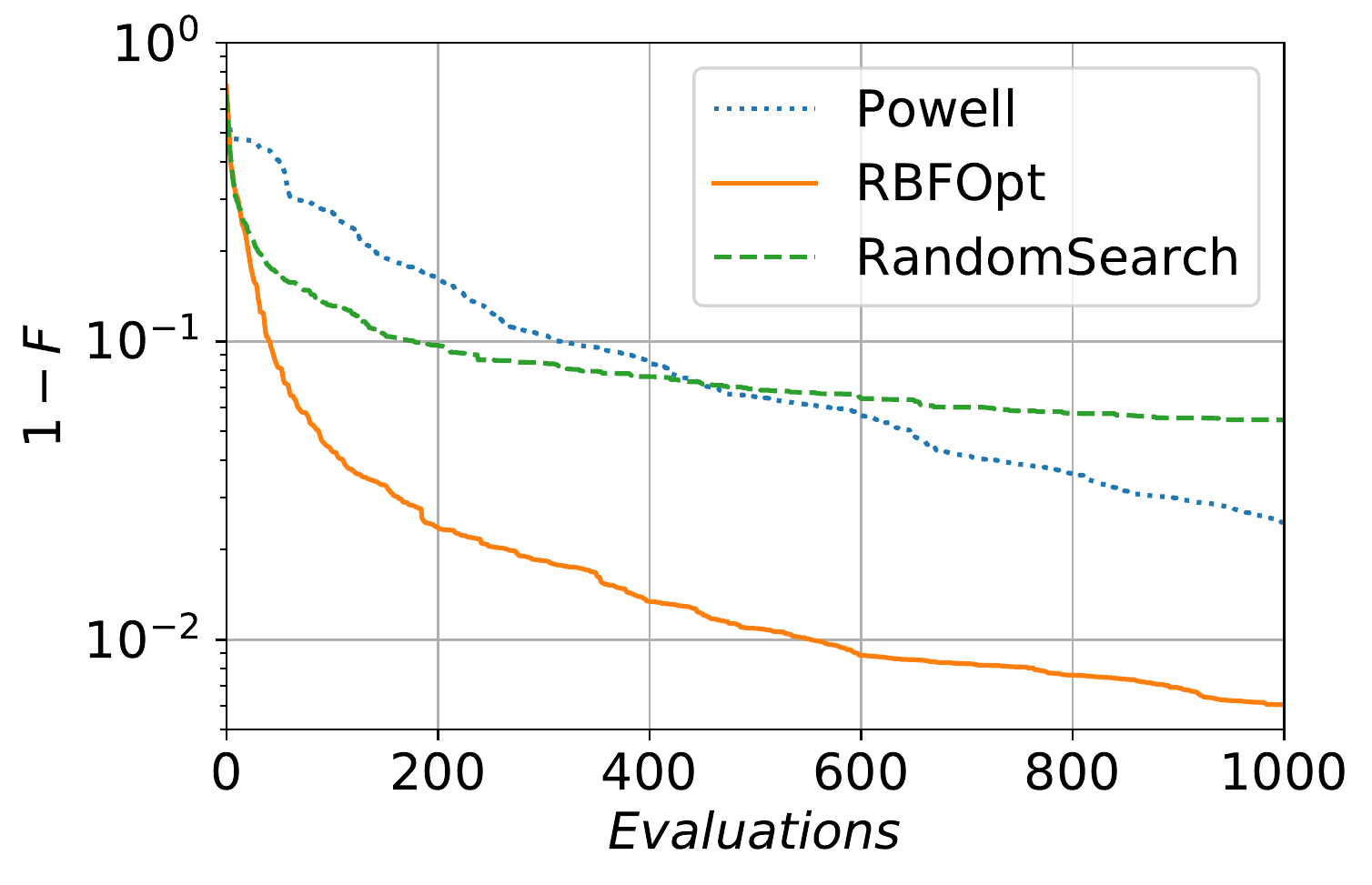}
\caption{\textbf{ Comparison  between  different  optimization algorithms:} The plot reports the simulated performances of three different algorithms averaged over the optimization of $10$ different states each of which is repeated $10$ times.
Dotted blue, dashed green, and continuous orange 
lines report the trends corresponding to Powell, Random Search, and RBFOpt, respectively.
RBFOpt is found to perform significantly better than the alternatives in most cases.
All curves are generated simulating experimental noise with both Poissonian ($\lambda=10^4$) and Binomial fluctuations. }

\label{Fig:Comp}
\end{center}

\end{figure*}

Regarding the first class, among the simplest there is the Random Search method. As suggested by the name, in each iteration of the optimization process the parameters are randomly extracted with a uniform distribution in the parameter space and independently from values assumed in previous steps.
The second comparative algorithm is based among simplest gradient-free adaptive methods known as the Powell method \cite{Powell_1964}. It attempts to find the local minimum nearest to the starting point. Initially, a set of directions is defined and the algorithm moves along one of them until a minimum is reached. This minimum becomes the uploaded starting point for the following minimization performed on the second direction. After repeating this procedure for each direction, a new direction is defined and the algorithm proceeds uploading the set of directions.  

In \cref{Fig:Comp} are reported the trends correspondent to each compared algorithm obtained averaging over the optimizations of $10$ distinct states each of which is repeated $10$ times. The experimental conditions are simulated adding both Poissonian ($\lambda=10^4$) and Binomial fluctuations. As expected, both the adaptive approaches result advantageous with respect to the random approach for a considerable number of function evaluations. Moreover, since the RBFOpt spans the whole parameter space through the global steps, its performances are substantially better.

\end{document}